# On the aftershocks of the Great Sumatra-Andaman earthquake

A. V. Guglielmi[1], O. D. Zotov[2], A. D. Zavyalov[1]

[1] *Schmidt Institute of Physics of the Earth, Russian Academy of Sciences, Moscow, Russia*

[2] *Borok Geophysical Observatory, Schmidt Institute of Physics of the Earth, Russian Academy of Sciences, Borok, Yaroslavl oblast, Russia*

E-mail: guglielmi@mail.ru, ozotov@inbox.ru, zavyalov@ifz.ru

**Abstract**

Analysis of the Sumatra-Andaman earthquake on 26.12.2004 (M = 9) has allowed us to identify two non-trivial properties of the dynamics of aftershocks. First, the strongest aftershock (M = 7.2) was likely triggered by the round-the-world seismic echo of the main shock. The idea is that the surface waves propagating outwards from the main shock return back to the vicinity of the epicenter after having made a complete revolution around the Earth and induce there the aftershock. The second property is the modulation of the aftershock sequence by the fundamental oscillation of the Earth $_0S_2$ excited by the main shock. Both results are supported by analysis of the Tohoku earthquake (11.03.2011, M = 9), as well as by the statistical analysis of the USGS earthquake catalog.

Contents

1. Introduction

An earthquake can be attributed to the area of critical phenomena. So it is natural to use a number of concepts and ideas from the theory of non-equilibrium dynamical systems and general catastrophes theory [Gilmore, 1981; Horsthemke, Lefever, 1984] when we are analysing the seismic data. In this work, we were guided by two well-known ideas concerning the universal properties of critical phenomena. The first one is that the amplitude of fluctuations increases with approaching the bifurcation point, so that at some moment a fairly strong internal pulse provides the critical transition (catastrophe). This transition will be called spontaneous. The second property is that sufficiently close to bifurcation the susceptibility of dynamical system dramatically increases. This means that even a weak external perturbation can cause the catastrophe. Critical phenomenon of this kind is naturally called the induced transition.

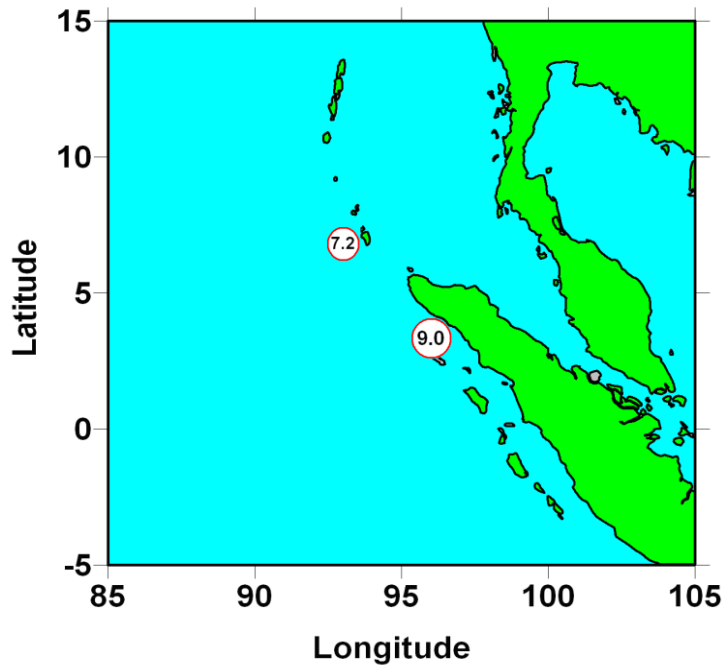

Fig. 1. The map of the North-Eastern region of the Indian Ocean. The epicenter of the main shock of the Sumatra-Andaman earthquake (M = 9) and the epicenter of the strongest aftershock (M = 7.2) are marked by circles.

In this paper, we use this knowledge in the analysis of aftershocks after the Sumatra-Andaman earthquake on 26.12.2004, M = 9 (e.g., see [Lay et al., 2005; Park et al., 2005; Zavyalov, 2005]). We paid attention to the two interesting manifestations of the induced seismicity. First is the large aftershock (M = 7.2) with a time delay of about 3 h 20 min respect to the main shock (Fig. 1). The idea is that the surface waves propagating outwards from the main shock return back to the vicinity of the epicenter after having made a complete revolution around the Earth and, in principle, may induce there an aftershock. The second manifestation is the modulation of the aftershock sequence by the fundamental oscillation of the Earth $_0S_2$ excited by the main shock.

In Section 2, which has a methodical character, we give an idea of endogenous and exogenous triggers of the earthquakes on the example of a simple dynamical system in a metastable state. In Sections 3 and 4 we present the arguments in favor of the hypothesis of seismic round-the-world echo and spheroidal oscillations of the Earth as the exogenous triggers that stimulate the aftershocks. The Sections 5 and 6 are devoted to discussion and conclusions respectively.

## 2. Phenomenology

### 2.1 Triggers of the catastrophes

In order to introduce the concept of endogenous and exogenous triggers, we use a simple phenomenological model that simulates the metastable state of the dynamical system. Such models are widely used, for example, in a qualitative analysis of critical phenomena in the magnetosphere [Guglielmi, Pokhotelov, 1996; Kangas et al., 1998]. Apparently, with certain reservations we can use a similar model in the discussion of critical phenomena in the lithosphere.

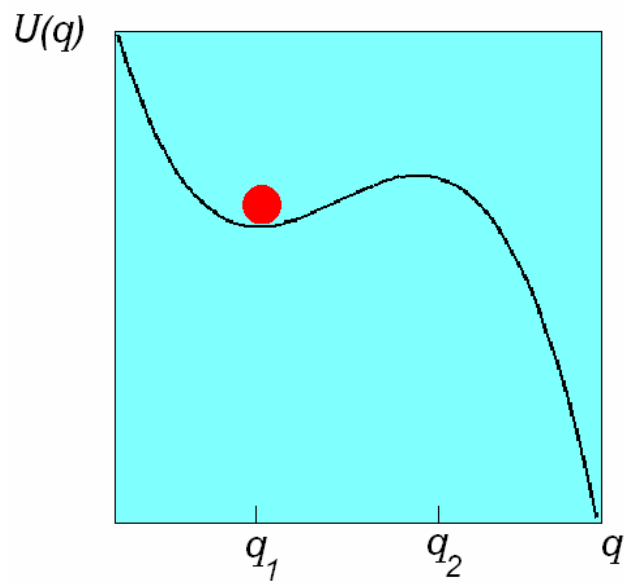

Fig. 2. The characteristic shape of the potential relief. Red ball is in a metastable state. Critical transition may occur under the action of endogenous or exogenous trigger.

Let $q(t)$ is the state of the system, the evolution of which in the absence of triggers is described by the equation

$$\frac{dq}{dt} = -\frac{\partial U}{\partial q}.$$  (1)

Here $U(q) = U(0) - A_1 q + A_2 q^2 - A_3 q^3$ is the effective potential, with $q \geq 0$, $A_i \geq 0$, $i = 1, 2, 3$. In other words, the potential has the form of a cubic parabola (Fig. 2). The minimum and maximum of the potential relief correspond to the stable ($q_1$) and unstable ($q_2$) equilibrium of the system ($dq/dt = 0$). State $q_1$ is metastable. This means that under the influence of internal noise or under an external impact the system can make a phase transition $q_1 \to q > q_2$ and thereby loses balance ($dq/dt > 0$).

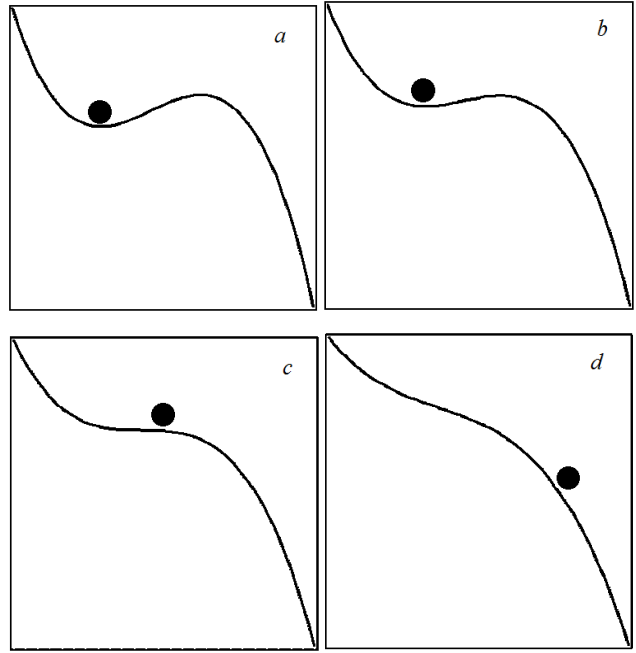

Fig. 3. Variation with the time of the effective potential $U(q)$ leading to disaster. From the metastable state (*a*, *b*) through a bifurcation (*c*) the system goes into a non-equilibrium state (*d*).

The system may lose balance without the influence of noise or external impact, if phenomenological parameters $A_i$ vary slowly so that the potential barrier height $\Delta U = U(q_2) - U(q_1)$ decreases monotonically. Fig. 3 illustrates this trend. Panel *c* shows the shape of the potential at the moment of critical transition from a metastable state of the system (panels *a* and *b*) to the non-equilibrium state (panel *d*).

The proposed model can be used as a kind of auxiliary scheme, but this should be done with caution. For example, if we ignore too obvious crudeness of our model and try to use it to describe the process of preparing an earthquake, it would seem, the moment of critical transition we need to consider as the moment of the earthquake. But it would be a mistake. Indeed, the background of seismic fluctuations exists always in the future focus of the forthcoming earthquake. Under their influence earthquake could occur before the critical transition, i.e. at the moment, as shown in panel *b*, rather than the panel *c* (see Fig. 3). In other words, in the vicinity of the bifurcation point, which is determined by the condition $\Delta U = 0$, some fluctuation of the stress field could be the trigger that causes an earthquake. The triggers of such sorts are naturally called endogenous. To describe them, we use a stochastic Langevin equation (2) instead of the dynamic equation (1):

$$\frac{dq}{dt} = -\frac{\partial U}{\partial q} + \xi(t) \tag{2}$$

Here the additive term $\xi(t)$ is a random function with zero mean, $\langle \xi(t')\xi(t'') \rangle = 2D\delta(t'-t'')$, where $\delta(t)$ is the Dirac delta function, and the angle brackets denote statistical averaging. A new phenomenological parameter $D$ is proportional to the intensity of the seismic noise in the upcoming earthquake focus. Similarly, we can take into account the endogenous triggers in the form of multiplicative noise, but we will not dwell on this (see for example [Horsthemke, Lefever, 1984]).

To further improve the model let us take into account external forces $F(t)$ acting on the system:

$$\frac{dq}{dt} = -\frac{\partial U}{\partial q} + \xi(t) + F(t). \tag{3}$$

Closer to the bifurcation of the barrier height $\Delta U$ decreases monotonically. Consequently, the reactivity of the dynamic system increases sharply near the threshold. In this state, even a weak external perturbation can cause a catastrophe. Such a critical transition is naturally called induced, and the corresponding trigger $F(t)$ is naturally called exogenous. If $F = 0$, the probability of transition is proportional to $\exp(-\Delta U / D)$, i.e. it is exponentially small for sufficiently high potential barrier [Kramers, 1940]. We note that if $F \neq 0$, the probability of transition can rise dramatically even with a relatively small amplitude of the external influence. The model (3) points to the increase in the probability of transition by the factor $\exp[(F/D)(q_2 - q_1)]$ (see the paper [Smelynskiy et al., 1999], which is devoted to the study of the important special case of a sinusoidal force $F$).

## 2.2 Two nontrivial triggers of the aftershocks

There is an extensive literature on the problem of earthquake excitation by the external forces. Exogenous triggers can be natural or artificial, pulsed or periodic; they can be terrestrial or cosmic origin (see for example [Nikolaev, Vereschagina, 1991; Hill et al., 1993, 2002; Hayakawa, 1999, 2012; Adushkin, Turuntaev, 2005; Zotov, 2007; Guglielmi, Zotov, 2012]).

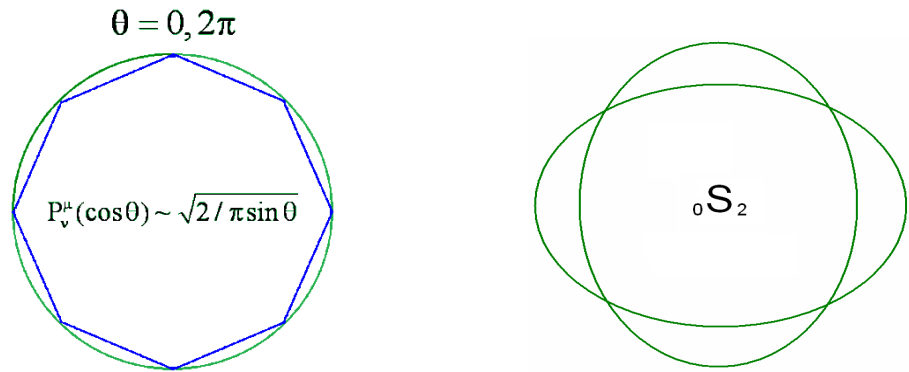

Fig. 4. Schematic pictures of the round-the-world seismic echo (left), and the spheroidal oscillations of the Earth (right).

We will focus on two exogenous triggers, the distinctive feature of which is that they originated as a result of the main shock of the Sumatra-Andaman earthquake, and presumably they influenced the dynamics of aftershocks. These are the round-the-world seismic echo, and the free oscillations of the Earth (Fig. 4). It is important to emphasize once again that both triggers are excited by the main shock of the earthquake.

The idea of the round-the-world echo is physically quite transparent. In highly simplified form, just look at the Earth as a specific lens, in which the source and the focus are in one place. The surface seismic waves are excited by the main shock, and are returned to the epicenter. The body waves can also create a round-the-world echo (see the blue line on the left panel of Fig. 4; it illustrates the resonant ray of whispering gallery type). The amplitude of the waves increases with the approach to the epicenter, because the epicenter is the focus (caustic). We believe that the round-the-world echo is capable of inducing a strong aftershock, since the crust in the vicinity of the epicenter is in the stress-strain state for a long time after the main shock.

It is well known that the earthquakes excite the free oscillations of the Earth as a whole at the resonant frequencies of toroidal and spheroidal eigen oscillations. This fundamental problem is investigated thoroughly both theoretically and experimentally (e.g., see the well-known books [Bullen, 1975; Aki, Richards, 1980; Zharkov, 1986, 2012]). Below we would like to present the arguments in favor of the idea that a reverse process also takes place. Namely, the Earth's free oscillations induce an earthquake activity. We will focus on the spheroidal oscillations $_0S_2$ (see Fig. 4, right panel), whose period is 54 min. We will present the evidence that the seismic activity is modulated with this period.

### 3. Round-the-world seismic echo

An attempt to find the trigger-effects in a series of strong aftershocks after the Sumatra-Andaman earthquake is justified by two considerations. First, the occurrence of numerous aftershocks suggests that stress level in the Earth's crust in the vicinity of the epicenter remained high for a long time after the main shock. The mainline rupture of the main shock did not remove the previously accumulated stress. It has redistributed them to other parts of the source zone, thereby increasing the likelihood of aftershocks. Therefore, it is reasonably to search for the induced earthquakes at a series of aftershocks above all. The second reason is less obvious. It is based on the hypothesis that exogenous triggers of aftershocks are formed in the earth's crust as a result of the main shock. In this section, we discuss the pulse trigger, which probably induced the strongest aftershock with a magnitude of M = 7.2, and in the next section, we consider the periodic trigger, the effect of which results in modulation of weaker aftershocks.

We suppose that the pulsed trigger originated in the form of the round-the-world echo. The Sumatra-Andaman earthquake has generated a surface elastic waves, which propagated from the epicenter with a characteristic rate of 3.7 km/s, and, having made a complete revolution around the Earth, returned to the epicenter at 3 h after the main shock. The amplitude of the surface wave has steadily intensified with the approach to the epicenter. The surface waves have intensified with the approach to the epicenter. Indeed, the Legendre function $P_\nu^\mu(\cos\theta)$ is proportional to the amplitude of the oscillations at an angular distance $\theta$ from the epicenter. Asymptotically $P_\nu^\mu(\cos\theta) \propto \sqrt{2/\pi\sin\theta}$ [Gradshteyn, Ryzhik, 1965], i.e., the amplitude of the round-the-world echo was growing with the approach to the epicenter $\theta = 2\pi$ (see Fig. 4, left panel). It is quite clear that the amplitude does not escalated to infinity, as indicated by the asymptotic theory in the framework of a spherically symmetric model of the Earth. An amplitude limitation occurred due to diffraction and also due to spherical and chromatic aberration. However, it can be assumed that the trigger came to the epicentral area in the form of a sufficiently powerful wave front approximately 3 hours after main shock of the Sumatra-Andaman earthquake.

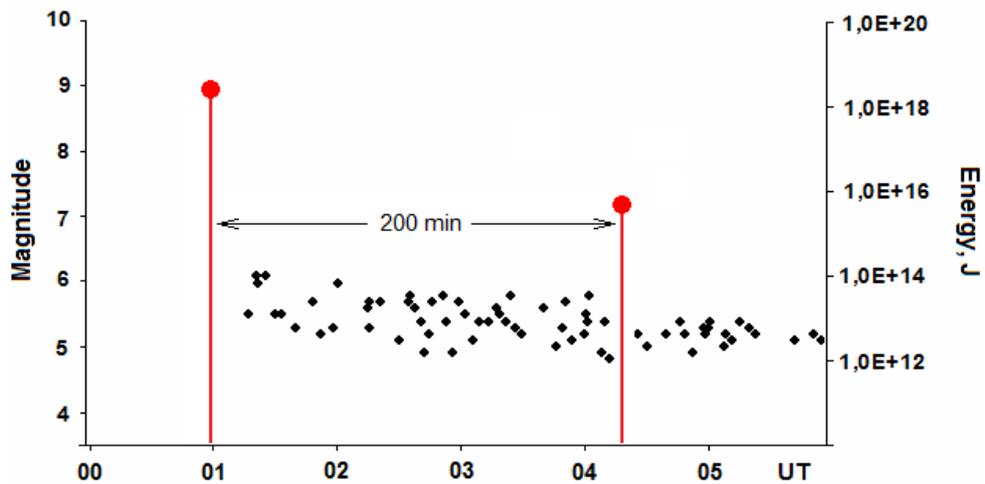

Fig. 5. Sumatra-Andaman earthquake of 26.12.2004. The red circles indicate the main shock and the strongest aftershock.

Let us turn, however, to the observations. Fig. 5 shows the aftershocks in the epicentral area of the radius $10^o$. To construct the figure we used data from the catalog of the National Earthquake Information Center of the US Geological Survey (NEIC, http://neic.usgs.gov/neis/epic/epic_global.html). We see that 70 aftershocks were recorded in the range of 5 h after the main shock. The strongest aftershock with M = 7.2 appeared with the delay of 3 h 20 min on the time of main shock. It is possible that it was induced by the seismic round-the-world echo. In this relation, it would be tempting to follow Anatole Abragam (1989) and speak of "a divine surprise of seeing the predicted phenomenon appear when it was expected, and such as it was expected", if not for one circumstance. The point is that we observe a difference of 20 min between the expected and the recorded time delay. If our interpretation is correct, the difference of 20 min is naturally explained by the phenomenon of after-effect ("overshot"), which is characteristic of the reaction of nonlinear dynamical systems to external stimuli.

## 4. Spheroidal oscillations of the Earth

As it was mentioned in the Section 2, the probability of transition in the dynamical system (3) increases sharply under the action of the periodic external force, even if amplitude of force is relatively small [Smelynskiy et al., 1999]. With regard to the Sumatra-Andaman earthquake, the periodic trigger naturally associates with the resonant oscillations of the Earth that have been generated by the main shock. Let us try to find the resonant oscillations in the spectrum of the sequence of aftershocks. We consider here the resonant frequency of the fundamental mode $_0S_2$ spheroidal oscillations (see Fig. 4, right panel). It is equal to 0.309 MHz, which corresponds to the period of 54 min [Aki, Richards, 1980; Zharkov, 1986, 2012].

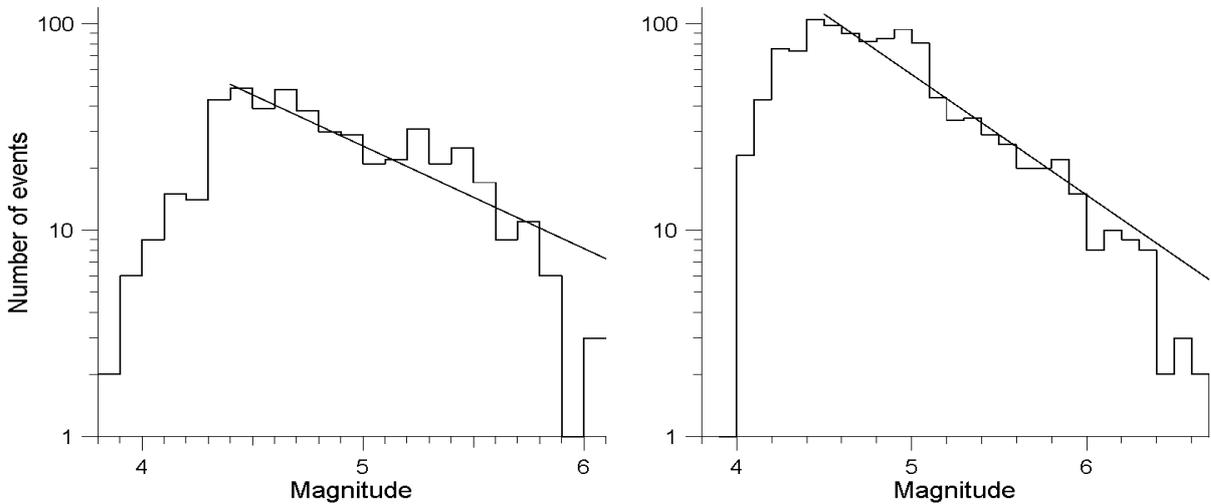

Fig. 6. Magnitude distributions of aftershocks of the Sumatra-Andaman earthquake (left) and the Tohoku earthquake (right) over the intervals 50 hours after the main shock in the epicentral zones of radius of 10°. Straight lines relate to the representative parts of the corresponding catalogues.

Initially, we select all significant aftershocks from the USGS catalog in the interval 50 hours after the Sumatra-Andaman earthquake in the epicentral zone of radius 10°. Fig. 6 on the left shows the distribution of aftershocks in magnitude M. The straight line $\lg N = 4.0 - 0.5M$ approximates a significant portion that corresponds to the magnitudes M > 4.4 and contains 357 events. Here $N$ is the number of aftershocks. The correlation coefficient is 0.85 at M > 4.4. In the future, we will need a similar distribution for the Tohoku earthquake. It is shown in Fig. 6 on the right. The straight line $\lg N = 4.67 - 0.58M$ approximates a significant portion that corresponds to the magnitudes M > 4.4 and contains 720 events. Correlation coefficient is 0.916 in this case.

Let us perform a spectral analysis of the sequence aftershocks. We assign zero for each second of 50-hour interval if there was no earthquakes at this moment, or a positive integer $\nu_j$, if there was $\nu_j$ earthquakes with M > 4.4 in the epicentral zone. We represent the dynamics of earthquakes as a series

$$n(t) = \sum_{j=1}^{N} \nu_j \delta(t - t_j), \qquad (4)$$

in which $t_j$ is the beginning of second interval, which is assigned a number $\nu_j$, $N$ is the total number of such intervals, and $\delta(t)$ is the Dirac delta function. Let us represent the function $n(t)$ as a Fourier integral:

$$n(t) = \int_{-\infty}^{\infty} n_\omega \exp(-i\omega t) \frac{d\omega}{2\pi} . \qquad (5)$$

Here the spectral component $n_\omega$ is given by the expression

$$n_\omega = \int_{-\infty}^{\infty} n(t) \exp(i\omega t) dt . \qquad (6)$$

Substituting (4) into (6), we find

$$n_\omega = \sum_{j=1}^{N} \nu_j \exp(i\omega t_j) . \qquad (7)$$

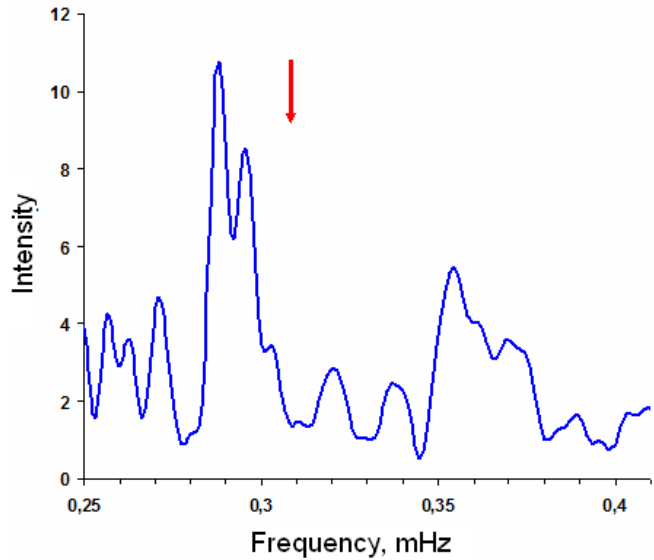

Fig. 7. The spectrum of aftershock activity over the interval of 50 hours after the Sumatra-Andaman earthquake. The arrow indicates the eigen frequency of the fundamental mode $_0S_2$ of the Earth's oscillations.

Fig. 7 shows the frequency dependence of the intensity of Fourier components $|n_\omega|^2$. The arrow indicates the frequency of spheroidal oscillations $_0S_2$. We see a significant increase of intensity in the band 0.28-0.3 mHz. Centre of the band deviates from the frequency of spheroidal oscillation by only a few percent. This suggests that the aftershock activity is apparently modulated by the spheroidal oscillations of the Earth. The above specified deviation may be due to the inaccuracy of the estimates of the intensity of the spectral components, calculated from the relatively short implementation earthquake sequence, or due to the omission of events in a powerful stream of aftershocks.

## 5. Discussion

### 5.1 Comparison with Tohoku earthquake

In Section 2, we pointed out the theoretical arguments in favor of the hypothesis of the existence of two non-trivial triggers that affect the activity of aftershocks. Analysis of the Sumatra-Andaman earthquake demonstrates the plausibility of our hypothesis. It is interesting to test our hypothesis additionally by using the data of the Tohoku earthquake (M = 9.0), which occurred March 11, 2011 at 05 h 46 min Greenwich Mean Time.

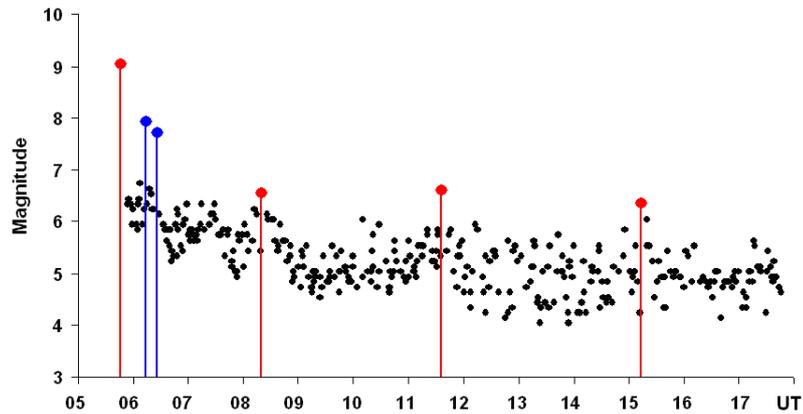

Fig. 8. Earthquake in the Tohoku region 11.03.2011 and aftershocks over the interval 12 hours after the main shock. Circles mark the main shock, and 5 of the strongest aftershocks.

In Fig. 8 we see the main shock and five strong aftershocks. Note the third, fourth and fifth aftershocks registered in 8 h 19 min, 11h 36 min, and 15 h 13 min with magnitudes M = 6.5, M = 6.6 and M = 6.3 respectively. The amazing regularity of their appearance suggests that we are dealing with earthquakes induced by the triple round-the-world seismic echo. As for the two powerful aftershocks that occurred at 6 h 16 min (M = 7.9) and 6 h 26 min (M = 7.7), it is likely that they were induced by the endogenous triggers.

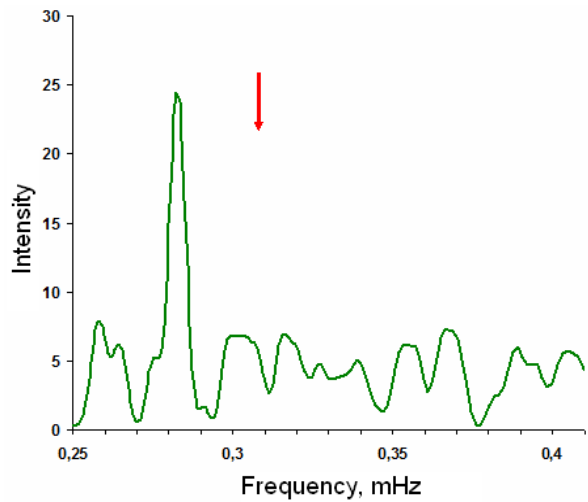

Fig. 9. The spectrum of aftershock activity over the interval of 50 hours after the Tohoku earthquake. The arrow indicates the eigen frequency of the fundamental mode $_0S_2$ of the Earth's oscillations.

Now let us consider the spectrum of aftershock activity after the Tohoku earthquake. The frequency dependence of intensity $|n_\omega|^2$ of the Fourier components is shown in Fig. 9. The arrow indicates the frequency of spheroidal oscillation $_0S_2$. The maximum at the frequency of 0.285 MHz deviates somewhat from the frequency of spheroidal oscillations, but this deviation may be due to the inaccuracy of estimates of the intensity of the spectral components. So, we have received another indirect confirmation of the hypothesis.

## 5.2 Statistical analysis

The analysis of particular events we supplement by the statistical analysis of a large number of earthquakes indicated in the catalog USGS. To test the hypothesis concerning the round-the-world seismic echo we used the method of superposition of epochs. Times of strong earthquakes are taken as the bench mark for synchronization of the aftershocks. Below we show the effect.

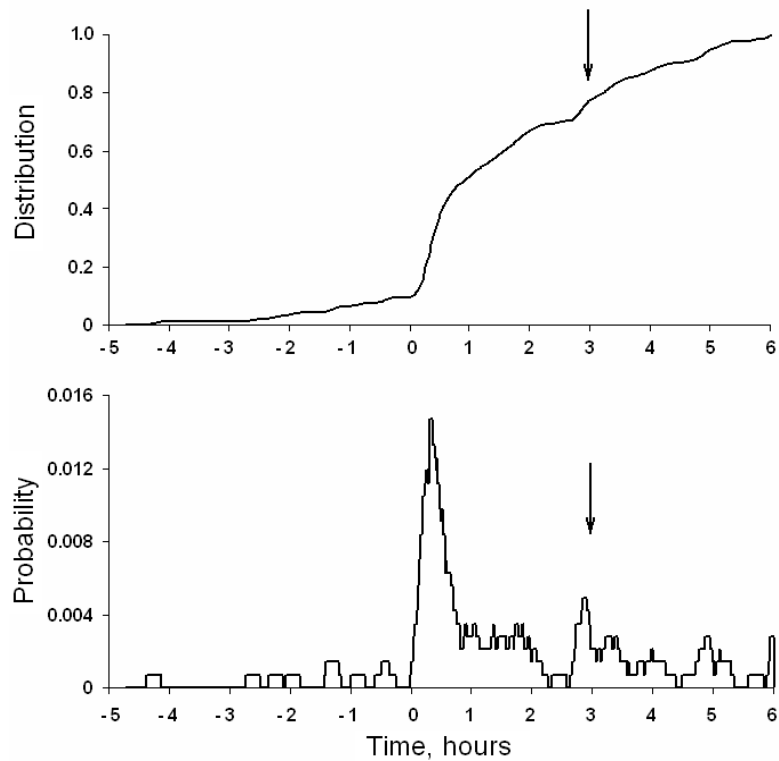

Fig. 10. Dynamics of the foreshocks and aftershocks in the epicentral zones of the strong earthquakes. The arrows indicate the expected time delay of round-the-world echo of the surface wave. The top panel shows the function of distribution of earthquakes in time relative to the reper. The bottom panel shows the density of probability of occurrence of an earthquake, depending on the time.

Fig. 10 shows the aftershock dynamics after 167 earthquakes with magnitudes $M \geq 7.5$ according to the USGS catalog from 1973 to 2010. We selected aftershocks with magnitudes $6 \leq M < 7.5$ in the epicentral zones with radius of 2°. The top panel shows the smoothed distribution function of earthquakes in time relative to the bench mark. The time dependence of the probability of earthquake occurrence is shown in the lower panel of Fig. 10. We see that the the most intense aftershock activity is observed within the first hour after the earthquake. After that, the aftershock activity decreases, and then begins the ascent, which is finished by a new maximum about 3 h after the main shock. It seems that our expectations have justified to some extent.

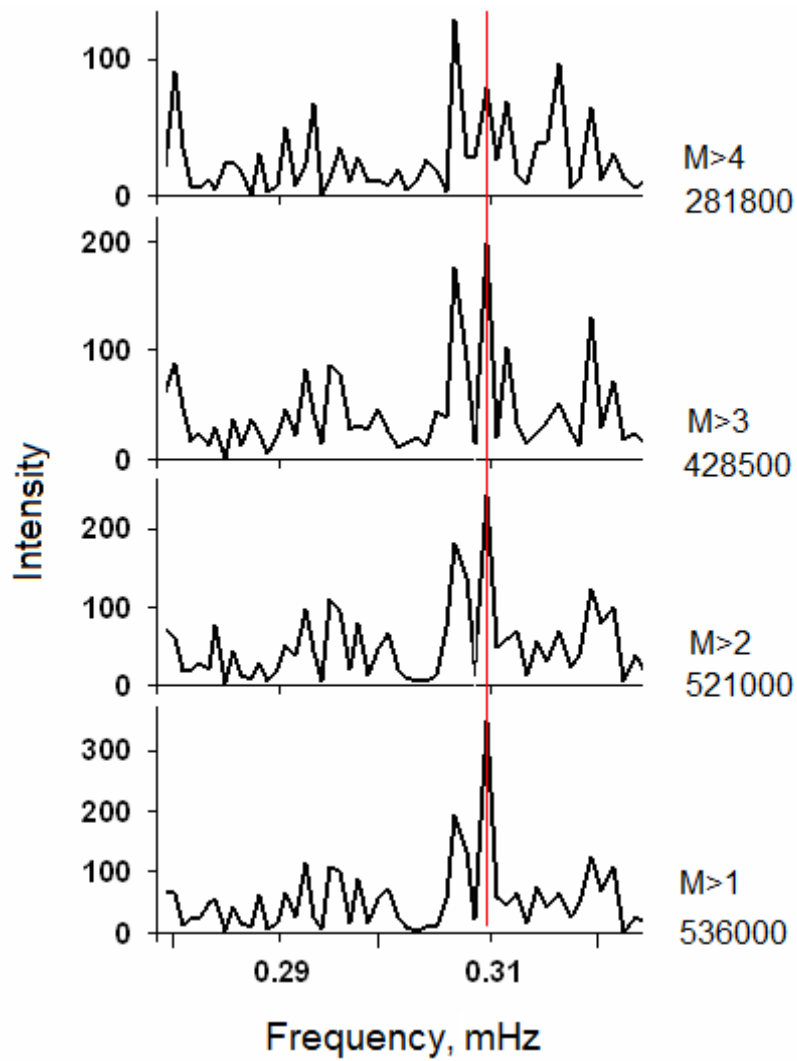

Fig. 11. Spectra of global seismic activity from 1973 to 2010. The minimum values of the magnitude M and the corresponding sample sizes are indicated to the right of each spectrum. The thin vertical line indicates the frequency of the fundamental mode of spheroidal oscillations of the Earth.

Concluding this section, let us consider the spectrum of global seismicity (Fig. 11). We divided the data of catalog USGS into four groups according to the minimum value of magnitude M. All the panels in Fig. 11 clearly demonstrate peaks at the frequencies of 0.307 mHz and 0.309 mHz. We believe that the Fig. 11 quite clearly shows the modulation of global seismicity under the action of spheroidal oscillations of the Earth.

## 6. Conclusion

The most important result of our work is that there are two exogenous triggers affecting the activity of aftershocks. The first of them is a pulse, and the second is periodical. The pulsed trigger is the round-the-world seismic echo, and the periodic trigger is the Earth's free oscillations. Both triggers are generated by the main shock of the earthquake.

The result is of interest for geophysics and, in particular, for the physics of induced seismicity. We hope that the result will be useful also in astroseismology, especially in the seismology of pulsars.

Toward the close of this paper, we would like to raise the question of whether or not to consider the global oscillations of the Earth as a self-oscillation. On the one hand, a positive answer to this question is obvious, as the Earth is largely autonomous dynamical system. On the other hand, it is not quite clear mechanism for self-excitation of oscillations per se. In particular, the feedback mechanism was previously not clear at all. Now we have some hope to understand the problem of feedback. Indeed, we see that the Earth's oscillations are excited by earthquakes, and at the same time these oscillations influence on the activity of earthquakes in some degree.

*Acknowledgments*. We thank the staff of USGS/NEIC for providing the catalogues of earthquakes. The work was supported by the RFBR (grants 12-05-00799 and 13-05-00066), and Programs of the Presidium RAS of Basic Research № 4 (Project 6.2), and Leading Scientific School of Russian Federation LSS-5583.2012.5.